\documentclass[iop]{emulateapj}

\usepackage{psfrag}

\usepackage{graphicx,shapepar}

\newcommand{\nii}{[\ion{N}{2}]}

\newcommand{\oiii}{[\ion{O}{3}]}

\begin{document}

\title{A Detailed Spatiokinematic Model of the Conical Outflow of the Multipolar Planetary Nebula, NGC 7026}

\author{Clark, D.~M. \altaffilmark{1}, L\'{o}pez, J.~A. \altaffilmark{1}, Steffen, W. \altaffilmark{1},  Richer, M.~G. \altaffilmark{1}}

\altaffiltext{1} {Instituto de Astronom\'{i}a, Universidad Nacional
  Aut\'{o}noma de M\'{e}xico, Campus Ensenada, Ensenada, Baja
  California, 22860; dmclark@astro.unam.mx}

\begin{abstract}

  We present an extensive, long-slit, high-resolution coverage
  of the complex planetary nebula (PN), NGC 7026.  We
  acquired ten spectra using the Manchester Echelle Spectrometer at San Pedro
  Martir Observatory in Baja California, Mexico, and each shows
  exquisite detail, revealing the intricate structure of this object.
  Incorporating these spectra into the 3-dimensional visualization and
  kinematic program, SHAPE, and using {\it HST} images of NGC 7026, we
  have produced a detailed structural and kinematic model of this PN.  NGC 7026 exhibits remarkable symmetry consisting of three lobe-pairs and four sets of knots, all symmetrical about the nucleus and displaying a conical outflow. Comparing the 3-D structure of this nebula to recent, {\it
    XMM-Newton} X-ray observations, we investigate the extended X-ray emission in relation to the nebular structure.  We find that 
  the X-ray emission, while confined to the closed, northern lobes of this
  PN, shows an abrupt termination in the middle of the SE lobe, which our long slit data shows to be open. This is where the shocked, fast wind seems to be escaping the interior of the nebula and  the X-ray emission rapidly cools in this region.
\end{abstract}


\section{Introduction}

NGC 7026 is a complex, bipolar planetary nebula (PN) as seen in images as well as high resolution spectra.  \citet{stan10} estimate a distance of 2086$\pm420$ pc  to the PN.  The central star is a [WC 3] star  \citep{koes01} with a V magnitude of 15.10 mag \citep{van95}.  It is a well studied PN that has received attention by many authors, in particular \citet{sol84}, \citet{cue96} and \citet{haj07}.  
\citet{sol84} obtained long-slit, image-tube coud\'e spectra.  They describe NGC 7026 as a bipolar configuration consisting of an expanding equatorial toroid and opposite expanding polar blobs. For the major axis they derive an inclination with-respect-to the line of sight  of 75$\degr$ and a position angle of 15$\degr$.    They notice that the nebula exhibits a strong ionization structure and describe NGC 7026 as an optically thin, moderately evolved PN. Their observations indicate a tendency of increasing expansion velocities with decreasing excitation. They also derive a distance for NGC 7026 of 2180 pc, in agreement with the more recent work quoted above. 

\citet{cue96} carried out a more detailed study of NGC 7026 using both high and low resolution spectroscopy and ground-based imaging.  They unveil a more complex morphological and kinematic structure and describe the PN as having four separate outflows or lobes with a central, spherical shell.  They derive  a mean electron density $n_e = 2.05 \times 10^3$ cm$^{-3}$ for the nebula and describe NGC 7026 as an evolving, bipolar PN still in its early stages of formation.

More recently,  \citet{haj07} have presented {\it HST} images and high resolution, echelle spectra for NGC 7026 centered on the lines of \nii{} 6584 \AA{}  and \oiii{} 5007 \AA{}. Their observations provide a very good spatial coverage across the nebula and yield detailed spectra of this intricate PN.  However, using models by \citet{aaq96} and \citet{zha98} the authors settle for a simple, bipolar model for this complex PN and therefore attain an unsatisfactory fitting of their synthetic spectra. Their data, though, are very valuable and they nicely complement the spectra presented in this work and together form a comprehensive source of long-slit, echelle spectra for NGC 7026.

Recently, \citet{gru06} acquired X-ray observations of NGC 7026 using
{\it XMM-Newton}.  The X-ray emission is confined within
the bipolar lobes and has a plasma temperature of T =
1.1$\times$10$^6$ K.  It is assumed that shock-heated gas is produced
  when the fast wind plows into the dense, slow wind of the AGB star.
  This should produce temperatures greater than 10$^7$ K, but
  generally the observed temperature is an order of magnitude lower
  \citep[see][and references therein]{geo06}.  \citet{geo06} suggest
  that the temperature drops through heat conduction between the hot
  gas and the cold, optical shell \citep[see also][]{sok94,zhe98}.
  Furthermore, mass evaporation raises the density of the shocked,
  fast wind.  Since the X-ray emission is associated with the
shocked fast wind, this emission should be a good indicator of {how the fast wind has been channeled through the nebula.

In this paper we use high resolution spectra, combined with the
spatio-kinematic program SHAPE, to understand the complex morphology
of this PN.  We organize this paper as follows: \S2 describes the
observations, \S3 discusses the results, \S4 discusses the SHAPE model and we finish with conclusions in \S5.

\section{Observations}

All observations of NGC 7026 were acquired on September 18, 2001,
at the Observatorio Astron\'{o}mico Nacional San Pedro M\'{a}rtir
(SPM), Baja California, M\'{e}xico.  We used the Manchester Echelle
Spectrometer (MES) with a SITe CCD detector on the 2.1 m
telescope.  This CCD consists of 1024$\times$1024 square pixels, each
24 $\mu$m wide.  All frames were binned two-by-two in both the spatial
and spectral directions, which yielded a spatial sampling of
0\farcs624 per bin.   The seeing variation during the observations averaged 1\farcs0 - 1\farcs8.   We used a 90 \AA{} bandwidth filter to
isolate the 87$^{th}$ order containing the H$\alpha$ and \nii{}
$\lambda\lambda$6548, 6584, nebular emission lines. Ten slit positions were obtained across the nebula, all of them oriented north - south. For nine of these
ten slit positions, we used a 70 $\mu$m wide slit, 5\farcm2 long.
This yielded a velocity resolution of 9.2 km s$^{-1}$ for the binning
we chose.  This spectral and spatial resolution is comparable to that reported by \citet{haj07}, 7.5 km s$^{-1}$ and 1\farcs0 - 1\farcs5 seeing; \citet{sol84}, 12 km s$^{-1}$ and 2\farcs0 seeing, and \citet{cue96}, 20 km s$^{-1}$ and 0\farcs9 - 1\farcs2.  In only one position, slit position g, we used a 150 $\mu$m wide slit, equivalent to a velocity resolution of 11.5 km s$^{-1}$. The MES-SPM spectra appear to be the deepest  of the sets mentioned above. The integration time for each pointing was 1800 s, except for slit g which had an exposure time of 1200 s. Each
observation was followed by a 200 s spectra of a Th/Ar lamp for
wavelength calibration.  We supplemented this
data set with a 60 s, H$\alpha$$+$\nii{} image.  Standard IRAF
routines were used during the data reduction process to correct for
bias, remove cosmic rays and to wavelength calibrate the spectra. The spectra were calibrated in
wavelength to an accuracy of $\pm 1$ km s$^{-1}$ when converted to radial velocity. All
spectra are corrected to heliocentric velocity ($V_{hel}$). These spectra are part of The SPM Kinematic Catalog of Galactic PNe \citep{lop12a} and are available at http://kincatpn.astrosen.unam.mx.

\begin{figure*}
\figurenum{1}
\center
\includegraphics[width=120mm]{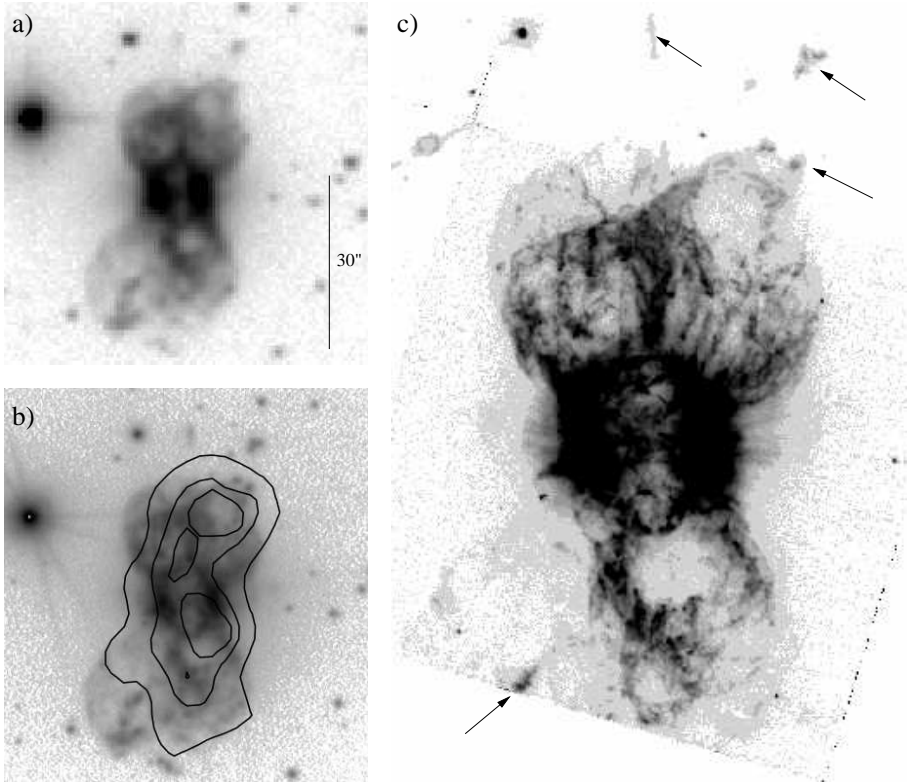}
\caption{Three different views of NGC 7026.  (a) MES-SPM,
  H$\alpha$$+$\nii{} image.  (b) Adapted from Figure 1 in \citet{gru06} by permission of the authors and reproduced by permission of the AAS, contours define extent of diffuse X-ray emission detected by {\it XMM-Newton} superimposed on a \nii{} NOT image.   (c)  {\it HST}  \nii{} image from the Hubble Legacy Archive, arrows indicate some of the knots discussed in the text. In all images north is up, east is to the left and a) shows the scale in arcseconds.
  \label{Fig.1}}
\end{figure*}

The SPM  H$\alpha$$+$\nii{} image is shown in Figure 1.
We show our observed slit positions overlaid on the MES-SPM
H$\alpha$$+$\nii{} image in Figure 2.  All positions were acquired
with the slit oriented N-S and each position is separated by
$\sim$3$\arcsec$, spread over the width of NGC 7026.  In this work, we
concentrated on the \nii{} $\lambda$6583 line profiles.  These
profiles provided the most detail, facilitating the modeling of this
PN.  All observed profiles are displayed in Figure 3 as
position-velocity (P-V) arrays, where we also include our model synthetic P-V
arrays produced using SHAPE (see below).

\section{Results and Discussion}

\subsection{Morphology}

In Figure 1, we present three different views of NGC 7026.  Panel a is an H$\alpha$$+$\nii{} image of NGC 7026 acquired with MES-SPM.  Panel b is taken from Figure 1 of  \citet{gru06} and shows the {\it XMM-Newton} X-ray emission as contours overlaid  on a NOT \nii{} image. The contour plots terminate abruptly half way through the south east lobe at a location where our spectra  shows (see section 3.3) that the lobe breaks open. Panel c is the {\it HST-PC} F658N image of NGC 7026 displayed in a logarithmic scale. This  {\it HST} image shows remarkable details of this intricate nebula. While the lobes can be seen in the MES-SPM image as fairly smooth structures, the {\it HST} image shows them to be formed by multiple filamentary loops with cometary knots  distributed along the inner edges of the lobes, pointing towards the central star. There are also emission knots beyond the extent of both lobes; given the long length of our slits, most of these knots have been detected in the spectra for the first time. Some of these knots are indicated in the image with arrows and are further discussed in the spectroscopy and modeling sections. The northern lobes appear slightly more compact than the southern counterparts, reaching a distance of 20$\arcsec$ from the central star, while the
southern lobes extend 27$\arcsec$ away from the nucleus.  Unfortunately the {\it HST} image does not cover the full extent of the southern region. The equatorial region is surrounded by a thick waist with radial emission spikes protruding to the outside and a knotty internal structure.

\subsection{Velocity Definitions}

Before continuing, we define the various velocity terminology that we use in the following sections.

$V_{hel}$ = Heliocentric velocity, it corresponds to the radial component of the space velocity vector along the line of sight  corrected for the earth's orbital motion, i.e. referred to the Sun. The velocity scale in the P -- V arrays presented in Figure 3 is $V_{hel}$. 

$V_{sys}$ = Systemic velocity, it corresponds to the mean $V_{hel}$ from the geometric center of the nebula (usually around the central star).  $V_{sys}$ is useful to describe expansion motions referred to the nebula's center, presumably the origin of any outflow.  

$V_{space}$ = Space velocity, it refers to the space velocity vector, obtained by correcting $V_{hel}$ for the tilt or inclination, $\theta$, of the major axis of an axisymmetric nebula with respect to the plane of the sky (or the line of sight).

  \begin{equation}
  V_{space}=V_{hel}/sin(\theta)
  \end{equation}

We use here slit f, which passes through the central star, to derive a
systemic velocity for NGC 7026 $V_{hel}$ = $-$41.4 km
s$^{-1}$.  This value is in agreement with those values derived by
\citet{sol84} of $V_{hel}$ = $-$40.5 km s$^{-1}$, by \citet{cam18} of
$V_{hel}$ = $-$40.3 km s$^{-1}$, and by \citet{sab82} of $V_{hel}$ =
$-$41.1 km s$^{-1}$. 

\subsection{Kinematics}

\begin{figure}
\figurenum{2}
\center
\includegraphics[width=85mm]{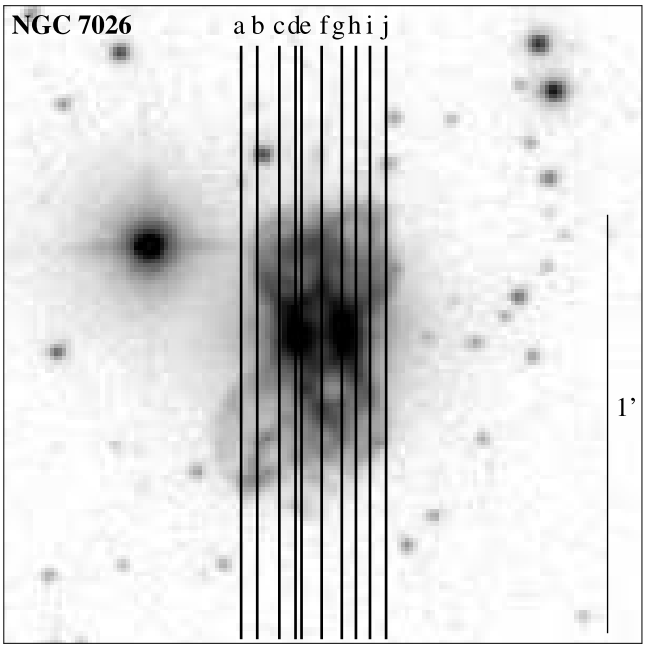}
\caption{H$\alpha$$+$\nii{}, SPM image with overlays of all slit
  positions acquired for NGC 7026. N is at the top and E is to the left.
\label{Fig.2}}
\end{figure}

\begin{figure*}
\figurenum{3}
\center
\includegraphics[width=118mm]{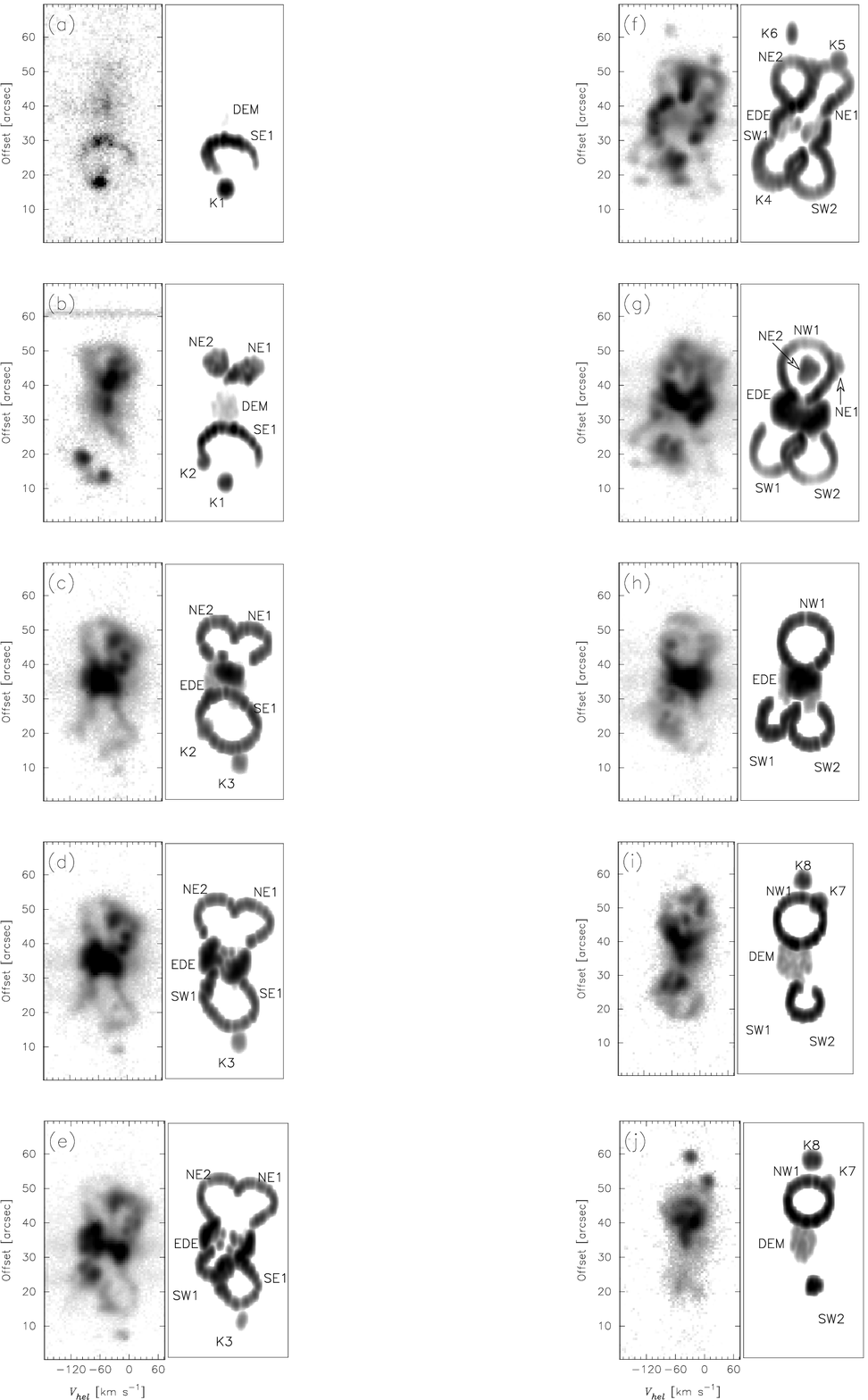}
\caption{P-V diagrams: observed and modeled.  Left panels
  are the observed \nii{} line profiles; right panels are the
  synthetic P-V line profiles produced by the SHAPE  model.  All major
  regions of NGC 7026 discussed in the text are labeled, knots K1--K8, lobes (NE1, NE2, NW1, SE1, SW1 and SW2),
  the EDE (Equatorial Density Enhancement) and the DEM (Diffuse Equatorial Material).
 \label{Fig.3}}
\end{figure*}

\subsubsection{Bipolar Lobes}

The main symmetry axis of NGC 7026 is projected nearly north - south (see Figure 2) and  slightly tilted with respect to the plane of the sky, as indicated by the P -- V array from slit f (see Figure 3), with the southern lobes pointed toward the observer and the northern ones away from the observer. For the main symmetry, taking the extreme opposite points along the middle of the line profile for slit f, we measure $V_{hel} - V_{sys} \simeq \pm 40$ km s$^{-1}$. An interesting feature of the observed P--V diagrams is the structure of the bipolar lobes.  From the long-slit spectra they look split on the NE (slits b -- f) and SW (slits f -- i) sections of the lobes, see Figure 3.  
We divided the lobes into four major sections, NW, SW, NE, and SE.  The SE section appears as one lobe as seen in slits a through d.  In slits a and b, the SE lobe appears open, while it is closed in slits c and d. In slit d, only a fraction of the lobe is covered.  The SW
 lobe seems pinched to form two structures, SW1 and SW2, as seen in slits f
  through i.  The northern lobes reflect in point symmetry the southern lobes.  In
  slits b through f, the NE lobes appear as two, NE1 and NE2, while
  the NW lobe appear as one in the remaining slits.

The placement and expansion velocities of the lobes suggest that they form a bi-conical structure.  We list the corresponding velocities for the lobes in Table 1.

A particularly interesting aspect of the lobes is the appearance of a
gap in the SE lobe.  This gap appears in the P-V diagrams for slit
positions a -- b.  The open nature of the SE lobe is also apparent in 
the G -- J spectra from \citet{haj07}.   As discussed in the introduction, X-ray
emission is usually confined to closed structures.  In the case of
this gap, there is a sharp termination of the X-ray emission \citep[Fig.1 of ][reproduced in  Figure 1, panel b, here]{gru06} at the location where the SE lobe seems open, indicating a quick thermalization or cooling of the hot, shocked gas in this region.

\begin{deluxetable}{lcc}
\tablecaption{Mean Heliocentric and Expansion Velocities for Lobes}
\tablewidth{200pt}
\startdata
\hline
\hline
Region & $V_{hel}$ & $V_{hel} - V_{sys}$\tablenotemark{a} \\
 & (km s$^{-1}$) & (km s$^{-1}$) \\
\hline
NE1 & 15.45 & 56.9 \\
NE2 & -70.90 & -29.5 \\
SE1 & -22.34 & 19.1 \\
NW1 & -1.11 & 40.3 \\
SW1 & -142.24 & -100.8 \\
SW2	& -12.32 & 29.1 \\
\enddata
\tablenotetext{a}{ $V_{sys}$ = $-$41.4 km s$^{-1}$.}
\end{deluxetable}

\begin{figure*}
\figurenum{4}
\center
\includegraphics[width=120mm]{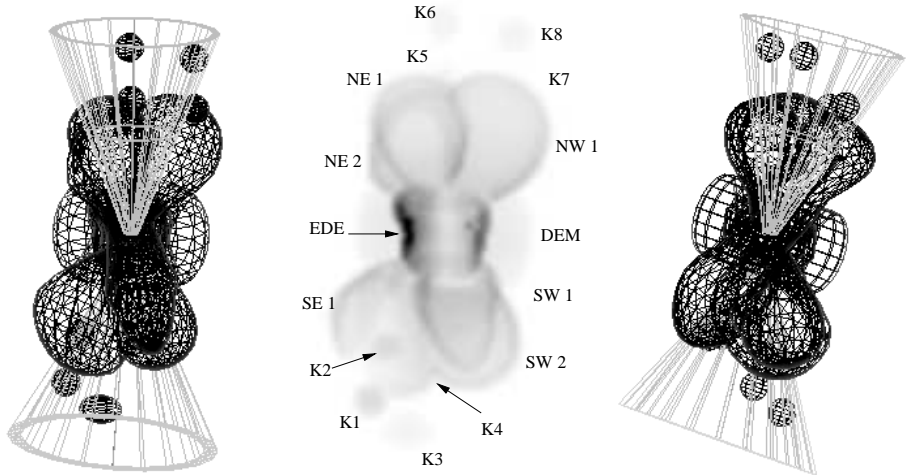}
\caption{ Left and right panels are wireframe views of  NGC 7026 where knots, lobes and the EDE are displayed together. A biconical surface that traces the location of the lobes and knots is also indicated. The left panel is a face-on view, i.e. as on the sky, and the right panel is a side view, rotated 90$\degr$  into the sky. The central panel shows the rendered model where individual knots and lobes have been labeled.
  \label{Fig.4}}
\end{figure*}

\subsubsection{Knots}

The spectra from all slit positions, except for g and h,  show compact knots of emission outside of the lobes or close to their borders. 
We found eight knots, four to the south and four knots to the north
of the nebula (see Figure 3). The heliocentric and expansion velocities for these knots are listed
in Table 2.

Interestingly, the knots appear to lie symmetrically about the central star and thus consisting of four pairs of knots: K1-K8, K2-K7, K3-K6, and K4-K5.  This association between knots is also reflected in the velocities, where each pair of knots have similar velocities, but opposite in sign (see Table 2).

\begin{deluxetable}{lcc}
\tablecaption{Heliocentric and Expansion Velocities for Knots}
\tablewidth{200pt}
\startdata
\hline
\hline
Region & $V_{hel}$ & $V_{hel} - V_{sys}$\tablenotemark{a} \\
 & (km s$^{-1}$) &  (km s$^{-1}$) \\
\hline
K1 & -52.7 &  -11.3  \\
K2 & -94.0 &  -52.6 \\
K3 & -15.7 &  25.7 \\
K4 & -125.5 &  -84.1 \\
K5 & 29.1 &  70.5 \\
K6 & -66.2 & -24.8 \\
K7 & 0.8 & 40.6 \\
K8 & -34.9 & 6.5 \\
\enddata
\tablenotetext{a}{$V_{sys}$ = $-$41.4 km s$^{-1}$.}
\end{deluxetable}

\subsubsection{Equatorial Region}

One of the most prominent characteristics in the P-V diagrams is the
fast expanding equatorial region we termed the equatorial density
enhancement (EDE).  In direct images of NGC 7026, the inner region
appears as a tight waist, but the P-V diagrams show the equator is expanding fast, nearly as fast as the lobes (see Figure 3, slit f).  Using slit f, we measured the peak heliocentric velocities from the front and back wall of the EDE to be $V_{hel}$=13.3 km s$^{-1}$ and $V_{hel}$=$-$101.3 km s$^{-1}$, respectively,  that translates into a direct expansion velocity $V_{exp}$=57.3 km s$^{-1}$.  Outside of the EDE is a region of diffuse emission
that can be seen in slits a, b, i, and j.  We labeled this region as
Diffuse Equatorial Material (DEM), which is nearly inert, i.e. with a velocity value close to the systemic. This is also the region where radial spikes are present in the {\it HST} image but show no kinematic counterpart, indicating photon leakage from the dense EDE, likely combined with scattering effects from surrounding warm dust in the core \citep{rob97}.

\section{SHAPE Model}

The high resolution spectra and close, angular separation of each slit
position across the nebula, makes these observations ideal for a model reconstruction using the
program SHAPE.  SHAPE \citep{ste06,ste11} is a tool that can be used to
obtain information on the three-dimensional structure
of a gaseous nebula from its kinematics and 2-D appearance on the sky. It requires
spatially resolved, high spectral resolution spectra with a good coverage over the nebula.  
The user can then use the
graphical interface to insert three-dimensional structures to
represent the form of the nebula.  These structures can be filled with
particles or the particles can be distributed across the surface.
Each system of particles can be given a unique velocity law.  In the
case of NGC 7026, we assumed a homologous expansion with a Hubble-type
 velocity law of the form $v = k
\cdot r/r_0$, where $k$ is a constant, $r$ is the distance from the
center, and $r_0$ is the distance at which the velocity $k$ is
reached.  When a desired structural representation of the nebula is
reached SHAPE then renders the system of particles at each slit
position, outputting synthetic spectra.  Through an iterative process
of changing the model and comparing the synthetic spectra to the
observed spectra, a three-dimensional form of the nebula can
be achieved \citep[see e.g.][]{gar09,cla10,lop12b}.

\begin{figure*}
\figurenum{5}
\includegraphics[width=175mm]{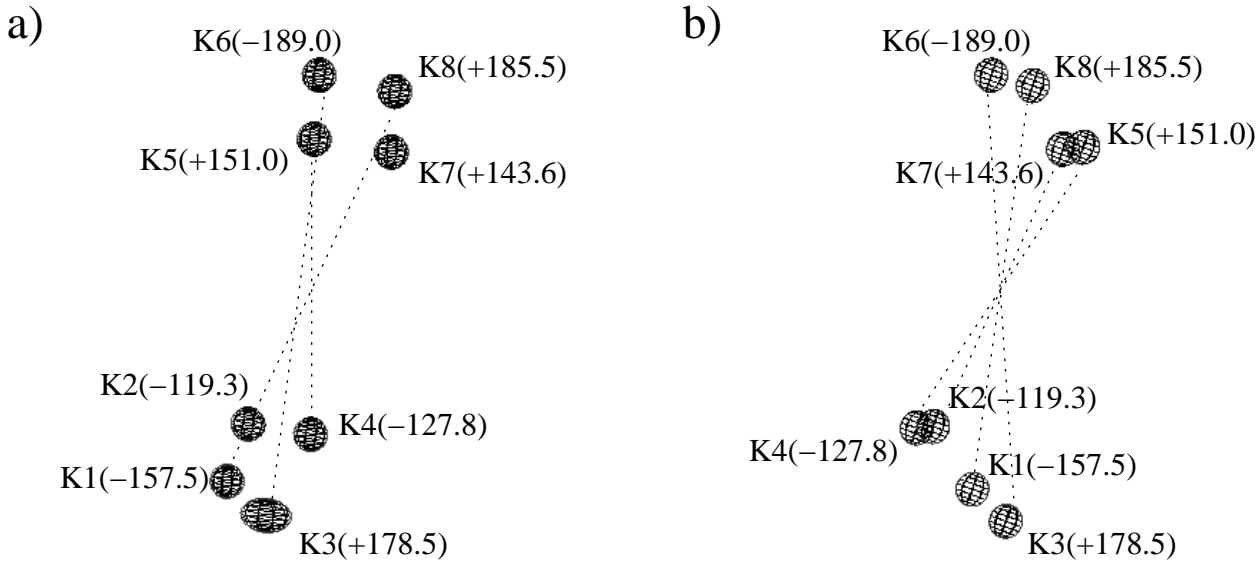}
\caption{Wireframe views of the system of point-symmetric knots. Panel a shows a face-on view, as see on the sky, panel b shows a side view, rotated 90\degr into the sky. The knots are labeled and dashed lines connect each pair. Numbers in parenthesis are the corresponding space velocities.
\label{Fig.5}}
\end{figure*}

We represented the lobed structure of NGC 7026 in our SHAPE model
using three, bipolar lobes.  These were grouped together to form the
lobes, which appear to split up as they grow out from the interior.  Particles were distributed across the surface of the lobes.    We modeled the gap in the SE lobe by setting the density to zero in the corresponding region with the form of a wedge.

The four sets of knots consist of eight, small spheres in our SHAPE
model.  Knot 3 extends across slits c--e,
so it was modified using an ellipsoid.  Close inspection of their locations and velocities indicate that the knots emanate from the lobes tracing different axes along a conical surface that opens up from the main nebular axis as they move away from the core.  
For the SHAPE model the knots were assigned the same homologous velocity law as the lobes and found to  approximately coincide as observed in projection, over the surface of two opposing conical sections.

Finally, we modeled the EDE using a tilted expanding toroid, with particles filling its
volume.

The main results of this modeling are shown in Figure 4.  The left and right panels show the full wireframe model, before rendering, where knots, lobes and EDE are displayed together; a biconical surface that follows the placement of knots and lobes is also indicated in these panels. The left panel shows a face-on, view, i.e. as seen on the sky and the right panel shows a side view, i.e. rotated 90$\degr$ into the sky; the central panel in this figure shows the rendered SHAPE model where knots and lobes are labeled. The synthetic P -- V derived from this model are shown next to the observed ones in Figure 3.

The main symmetry axis is found to be tilted 15$\degr$$\pm 2\degr$ with respect to the plane of the sky, considering the derived expansion velocity (see section 3.3.1)   $V_{hel} - V_{sys} \simeq \pm 40{}$ km s$^{-1}$, we find that the mean value for the deprojected velocity along the main symmetry axis of NGC 7026 is $V_{space}$ $\simeq{}\pm$150 km s$^{-1}$. Adopting a distance of 2000 pc to the nebula and a mean size for the lobes of 25\arcsec, the approximate expansion age of this PN is 1.54$\times$10$^{3}$ yr.

The knots and lobes open out from the main symmetry axis with distance from the core in a conical mode, and each of them presents a different tilt angle. As mentioned before, SHAPE allows the user to specify a velocity law for a system of particles.  Thus,  from our final model we can pick a region of the nebula and read it's space  or deprojected  velocity, irrespective of the inclination of the PN. Since the set of knots are localized, compact structures, it is relatively easy to identify  them and  obtain their space velocities from the final SHAPE model. This information is then combined with the derived expansion velocities to solve equation 1 for $\theta$. Table 3 lists the angles with respect to the plane of the sky for the trajectories over which each knot is traveling and their corresponding space velocities. Figure 5 shows two wireframe views of the system of knots, panel a shows a face-on, view, i.e. as seen on the sky and panel b shows a side view, i.e. rotated 90$\degr$ into the sky; the knots are labeled and dashed lines connect each pair. Numbers in parenthesis are the corresponding space velocities. Knots K 1 -- K8 and K3 -- K6 are the pairs located furthest away from the center of the nebula and also those with the largest velocities, $\approx 160 - 190$ km s$^{-1}$. The other knots are located close to the lobe borders and show outflowing space velocities, $\approx 120 - 150$ km s$^{-1}$, comparable to the main bipolar outflow.

\begin{deluxetable}{lcc}
\tablecaption{Space Velocities for the Knots}
\tablewidth{200pt}
\startdata
\hline
\hline
Region  & $\theta$\tablenotemark{a} & $V_{space}$ \\
 & ($\degr$) & (km s$^{-1}$) \\
\hline
K1 & -4.1 & -157.5 \\
K2 & -26.2 & -119.3 \\
K3 & 8.3 & 178.5 \\ 
K4 & -41.1 & -127.8 \\
K5 & 27.8 & 150.7 \\
K6 & -7.5 & -189.0\\
K7 & 16.4 & 143.6\\
K8 & 2.0 & 185.5\\
\enddata
\tablenotetext{a}{Angle with-respect-to the plane of the sky calculated using $V_{exp}$.  See text for details.}
\end{deluxetable}

\section{Conclusions}

In this work we studied the kinematics and 3-dimensional structure of
the PN NGC 7026, using high resolution spectra acquired with MES-SPM and a \nii  {}
{\it HST} image from the Hubble Legacy Archive. The {\it HST} image shows NGC 7026 to be formed by multiple filamentary loops with cometary knots distributed along the inner edges of the lobes. Several emission knots are found beyond the extent of the lobes. The region close to the core, enclosed by the equatorial density enhancement is filled with filaments and diffuse material.

We modeled the spectra using the program SHAPE to
interactively explore the structure and kinematics of this PN.  We
found that NGC 7026 is a poly-polar nebula, consisting of three entangled bipolar lobes, a
fast-expanding equatorial waist, diffuse material outside the waist,
and four pairs of high-speed knots of emission.  The main outline of the outflow, i.e. lobes and knots, has acquired a biconical mode that opens out as material moves away from the core.
These findings are in line with the previous interpretation of \citet{cue96} that found two pairs of bipolar lobes.

We find that the nebula is tilted by 15$\degr$ with respect to the plane of the sky, in agreement with  \citet{sol84}. Along the main symmetry axis the nebula is expanding at 150 km s$^{-1}$.  The estimated kinematic age for NGC 7026 is only 1.54$\times$10$^{3}$ yr.  Surrounding the central regions is a bright waist of emission produced by an equatorial density enhancement that is expanding at $\sim$57 km s$^{-1}$, such fast expansion of an EDE is uncommon, another example is NGC 6751 \citep{cla10}.  Outside of the EDE is a region of diffuse material that sits at the systemic velocity. The {\it HST} image shows here emission spikes apparently emerging from the EDE that seem to be produced by scattering effects from surrounding warm dust.  
The three sets of lobes show point symmetry in space and velocity. 
From our long-slit spectra we found that the SE lobe is open ended. The gap of optical emission in the bottom section of the SE lobe is also apparent in the spectra from \citet{haj07} and it coincides with  the  border of extended X-ray emission reported  by \citet{gru06} in this region. It seems  likely that the shocked wind thermalizes quickly when reaching the open-ended region of the SE lobe, bringing the X-ray emission to an abrupt termination. Beyond the extent of the lobes and close to the their borders we find eight emission knots that also  form four point-symmetric pairs, both in relation to their placement with respect to the core and in velocity. Some of these knots reach space velocities of the order of 180 km s$^{-1}$. These knots seem to be high-speed parcels of gas related to the early stages of formation of the poly-polar structure.

The overall scenario is that of a PN whose wind speed and ionization structure have developed fast in the recent past, producing shocks (extended X-ray emission) and hydrodynamic instabilities (cometary knots and filamentary lobes) in the surrounding environment, leading to a complex bipolar structure during its evolution. According to Koersteke (2001) the central star of NGC 7026 is  a [WC3] type star with an effective temperature T$\rm_{eff} = 130.5$ kK, terminal wind velocity V$\infty = 3500$ km s$^{-1}$ and a mass loss rate log $\dot{M} = - 6.34$ M$_{\odot}$ yr$^{-1}$. These parameters imply that the  stellar mechanical energy output rate is of the order of $2 \times 10^{36}$ erg s$^{-1}$. It is of interest to compare this energy budget with the one required by the outflowing nebular gas. The {\it HST} image shows that NGC 7026 is highly filamentary, with large and uncertain filling factors in the lobes. We shall assume a uniform average density of 2.05$\times 10^3$ cm$^{-3}$ \citep{cue96} and in order to calculate the mass contained in a full lobe we derive the volume for a frustum of a cone with the corresponding dimensions for NGC 7026, yielding 0.13 M$\odot$. For this amount of gas outflowing at velocities of 150 km s$^{-1}$ for 1.54 $\times 10^3$ years, a gas kinetic energy of 6.4 $\times 10^{35}$ erg s$^{-1}$ is required.  It is likely that this requirement is overestimated since we are considering a completely filled volume for the lobes, but it shows that the central star has ample mechanical power in its wind to drive the observed expansion patterns in the ionized gas presented in this work.

\section*{Acknowledgments}

This research has benefited from the financial support of DGAPA-UNAM
through grants IN116908, IN108506, IN100410, IN110011 and CONACYT 82066.
We acknowledge the excellent support of the technical personnel at the OAN-SPM, particularly Gustavo Melgoza, Felipe Montalvo and Salvador Monrroy, who were the telescope
operators during our observing runs. We thank R. Gruendl, M. A. Guerrero and Y-H Chu for their kind permission to use here their Figure 1c from Gruendl et al (2006). We also thank the anonymous referee whose constructive comments helped to improve the presentation of this work.


\clearpage

\end{document}